\documentclass{kluwer}    

\newcommand{\gapprox}{\lower.4ex\hbox{$\;\buildrel >\over{\scriptstyle\sim}\;$}}
\newcommand{\lapprox}{\lower.4ex\hbox{$\;\buildrel
<\over{\scriptstyle\sim}\;$}}

\usepackage{epsfig}

\begin{document}
\begin{article}
\begin{opening}
\title{Tomographic 3D-Modeling of the Solar Corona with FASR}

\author{Markus J. \surname{Aschwanden}, 
		David \surname{Alexander}, and
		Marc \surname{DeRosa} 		}

\runningauthor{ASCHWANDEN, ALEXANDER, \& DEROSA }

\runningtitle{FASR CORONAL TOMOGRAPHY}

\institute{Lockheed Martin Advanced Technology Center,
                Solar \& Astrophysics Laboratory,
                Dept. L9-41, Bldg.252,
                3251 Hanover St.,
                Palo Alto, CA 94304, USA;
                e-mail: aschwanden@lmsal.com }

\date{2002/Oct/14}

\begin{abstract}
The {\sl Frequency-Agile Solar Radiotelescope (FASR)} litteraly opens up a new
dimension in addition to the 3D Euclidian geometry: the frequency dimension.
The 3D geometry is degenerated to 2D in all images from astronomical telescopes,
but the additional frequency dimension allows us to retrieve the missing third
dimension by means of physical modeling. We call this type of 3D reconstruction
{\sl Frequency Tomography}. In this study we simulate a realistic 3D model
of an active region, composed of 500 coronal loops with the 3D geometry
$[x(s),y(s),z(s)]$ constrained by magnetic field extrapolations and the 
physical parameters of the density $n_e(s)$ and temperature $T_e(s)$ given
by hydrostatic solutions. We simulate a series of 20 radio images in a 
frequency range of $\nu=0.1-10$ GHz, anticipating the capabilities
of {\sl FASR}, and investigate what physical information can be retrieved 
from such a dataset. We discuss also forward-modeling of the chromospheric and
Quiet Sun density and temperature structure, another primary goal of future 
{\sl FASR} science.  
\end{abstract}


\end{opening}

\section{	INTRODUCTION                    }

Three-dimensional (3D) modeling of solar phenomena has always been a challenge
with the available two-dimensional (2D) images, but is an utmost necessity
to test physical models in a quantitative way. Since solar imaging telescopes
never have been launched on multiple spacecraft that separate to a significant
parallax angle from the Earth, no true 3D imaging or solar tomography (Davila 1994;
Gary, Davis, \& Moore 1998; Liewer et al. 2001) has been performed so far.
The {\sl Solar TErrestrial RElations Observatory (STEREO)}, now being assembled
and planned for launch in November 2005, will be the first true stereoscopic
facility, mapping the Sun with an increasing separation angle of $22^\circ$ per
year. Alternative approaches of 3D reconstruction methods utilize the solar 
rotation to vary the aspect angle (Altschuler 1979; Berton \& Sakurai 1985;
Koutchmy \& Molodensky 1992; Aschwanden \& Bastian 1994a,b; Batchelor 1994;
Hurlburt et al. 1994; Zidowitz 1999; Koutchmy, Merzlyakov, \& Molodensky 2001), but this
method generally requires static structures over several days. An advanced form of 
solar rotation stereoscopy is the so-called {\sl dynamic stereoscopy}
method (Aschwanden et al. 1999, 2000a), where the 3D geometry of dynamic plasma 
structures can be reconstructed as long as the guiding magnetic field is 
quasi-stationary. Of course, 3D modeling with 2D constraints can also be 
attempted if a-priori assumptions are made for the geometry, e.g. using the
assumption of coplanar and semi-circular loops 
(Nitta, VanDriel-Gestelyi, \& Harra-Murnion 1999). 

A new branch of 3D modeling is the combination of 2D images $I(x,y)$ with the 
frequency dimension $\nu$, which we call {\sl frequency tomography}. There
have been only very few attempts to apply this method to solar data, 
mainly because multi-frequency
imaging was not available or had insufficient spatial resolution. 
There are essentially only three published studies that employ the
method of frequency tomography: Aschwanden et al. (1995); 
Bogod \& Grebinskij (1997); Grebinskij et al. (2000.). 

In the first study (Aschwanden et al. 1995), gyroresonance emission above a 
sunspot was observed at 7 frequencies in both polarizations in the frequency 
range of $\nu=10-14$ GHz with the {\sl Owens Valley Radio Observatory (OVRO)} 
during 4 days. From stereoscopic correlations the height levels $h(\nu )$ 
of each frequency could be determined above the sunspot. Correcting for the
jump in height when dominant gyroresonance emission switches from the
second ($s=2$) to the third harmonic ($s=3$), the magnetic field 
$B(\nu )=357 ({\nu}_{GHz}/s)$ [G] could then be derived as a function of
height, $B(h)$, and was found to fit a classical dipole field
$B(h)=B_0 (1 + h/h_D)^{-3}$. Moreover, from the measured brightness
temperature spectrum $T_B(\nu )$, using the same stereoscopic height 
measurement $h(\nu )$, also the temperature profile $T(h)$ as a function
of height above the sunspot could be determined. This study represents 
an application of {\sl frequency tomography}, additionally supported 
with {\sl solar rotation stereoscopy}, and thus is subject to the requirement
of quasi-stationary structures.

In the second study (Bogod \& Grebinskij 1997), brightness temperature spectra
$T_B(\nu)$ were measured in 36 frequencies in the wavelength range of
$\lambda = 2-32$ cm ($\nu=0.94-15$ GHz) with RATAN-600,
 from quiet-Sun regions, active region
plages, and from coronal holes. A differential deconvolution method of
Laplace transform inversion was then used to infer the electron temperature
$T(\tau )$ as a function of the opacity $\tau$. This method does not yield
the temperature as a function of an absolute height $h$, but if an atmospheric
model [$T(h), n_e(h)$] is available as a function of height, the temperature 
as a function of the free-free (bremsstrahlung) opacity $T(\tau )$ can be 
calculated and compared with the observations. 

In the third study (Grebinskij et al. 2000), the brightness temperature
in both polarizations is measured as a function of frequency, i.e.
$T_B^{RCP}(\nu )$ and $T_B^{LCP}(\nu )$. Since the magnetic field has a
slightly different refractive index in the two circular polarizations, 
the free-free (bremsstrahlung) opacity is consequently also slightly 
different, so that the magnetic field $B(\nu )$ can be inferred.
Again, a physical model [$T(h), n_e(h), B(h)$] is needed to predict 
$B(\nu )$ and to compare it with the observed spectrum $T_B(\nu)$.

The content of this paper is as follows: In Section 2 we simulate 
an active region, with the 3D geometry constrained by an observed 
magnetogram and the physical parameters given by hydrostatic solutions, 
which are used to calculate FASR radio images in terms of brightness
temperature maps $T_B(x,y,\nu )$, and test how the physical parameters
of individual coronal loops can be retrieved with FASR tomography. 
In Section 3 we discuss a few examples of chromospheric and Quiet-Sun
corona modeling to illustrate the power and limitations of FASR
tomography. In the final Section 5 we summarize some primary goals
of FASR science that can be pursued with frequency tomography.

\section{ 	ACTIVE REGION MODELING	  }

\subsection{	Simulation of FASR Images }

\begin{figure}
\epsfig{file=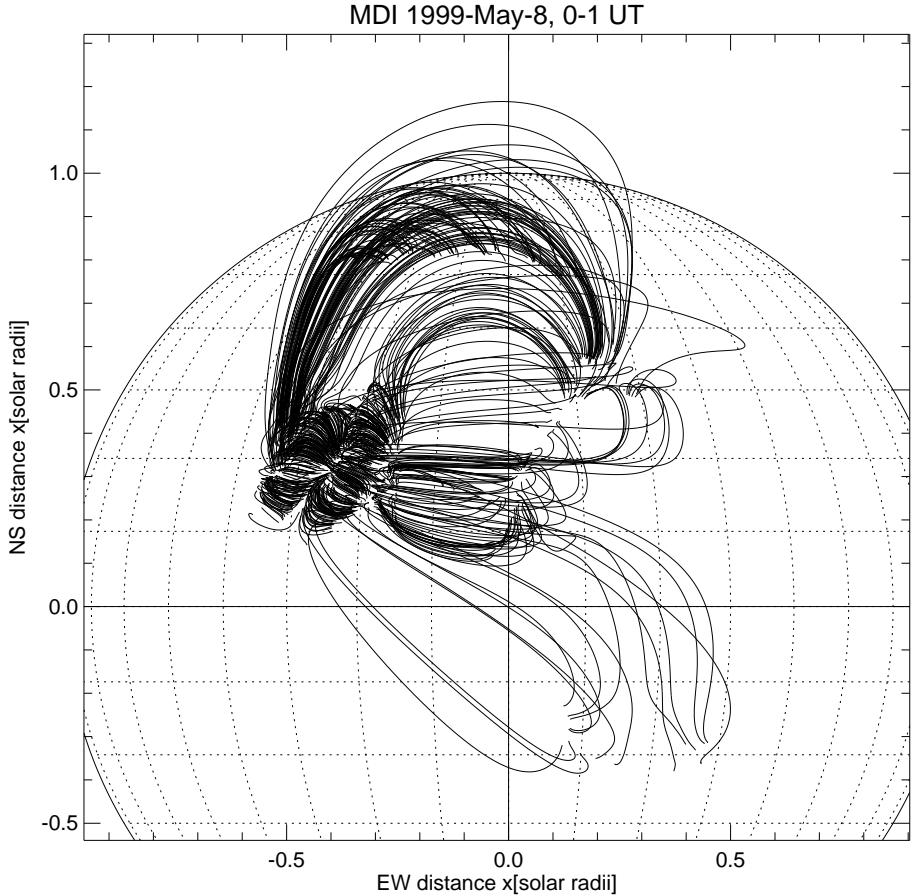,width=\textwidth} 
\caption{Potential field extrapolation of {\sl SoHO/MDI} magnetogram 
data from 1999 May 8, 0-1 UT.}
\end{figure}

Our aim is to build a realistic 3D model of an active region, in form
of 3D distributions of the electron density $n_e(x,y,z)$ and electron
temperature $T_e(x,y,z)$, which can be used to simulate radio brightness
temperature maps $T_B(x,y,\nu)$ at arbitrary frequencies $\nu$ that can
be obtained with the planned {\sl Frequency-Agile Solar Radiotelescope (FASR)}.

We start from a magnetogram recorded with the {\sl Michelson Doppler Imager
(MDI)} instrument onboard the {\sl Solar and Heliospheric Observatory (SoHO)}
on 1999 May 8, 0-1 UT. We perform a potential field extrapolation with the
magnetogram as lower boundary condition of the photospheric magnetic field,
to obtain the 3D geometry of magnetic field lines. We apply a threshold for
the minimum magnetic field at the footpoints, which limits the number of
extrapolated field lines to $n=500$. The projection of these 3D field lines
along the line-of-sight onto the solar disk is shown in Fig.1. We basically see
two groups of field lines, (1) a compact double arcade with low-lying field lines
in an active region in the north-east quadrant of the Sun, and (2) a set of
large-scale field lines that spread out from the eastern active region to 
the west and close in the western hemisphere. From this set of field lines we have
constrained the 3D geometry of 500 coronal loops, defined by a length coordinate
$s(x,y,z)$.

\begin{figure}
\epsfig{file=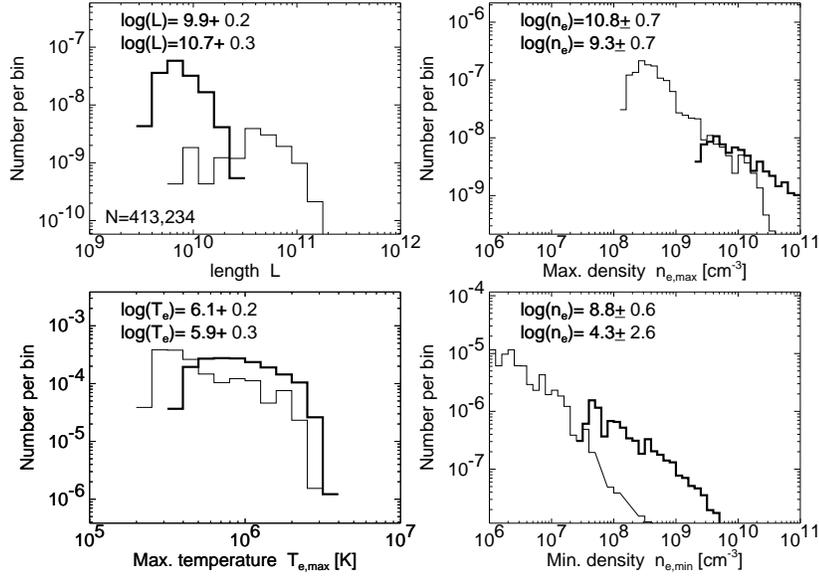,width=\textwidth} 
\caption{Distributions of loop lengths $L$, 
loop maximum temperatures $T_{e,max}$,
loop minimum densities $n_{min}$, and maximum densities $n_{max}$. The
distributions with thick linestyle correspond to $\approx 400$ loops in the
compact arcade, while the distributions with thin linestyle correspond
to the group of $\approx 100$ large-scale loops.}
\end{figure}

In a next step we fill the 500 loops with coronal plasma with density $n_e(s)$
and temperature functions $T_e(s)$ that obey hydrostatic solutions. For
accurate analytical approximations of hydrostatic solutions we used the code
given in Aschwanden \& Schrijver (2002). Each hydrostatic solution is defined
by three independent parameters: the loop length $L$, the loop base heating
rate $E_{H0}$, and the heating scale height $s_H$. The momentum and energy
balance equation (between the heating rate and radiative and conductive loss rate, i.e. $E_H(s)+E_{rad}(s)+E_{cond}(s)=0$, yields a unique solution for each parameter set $(L, s_H, E_{H0})$. For the set of short loops
located in the compact double arcade, which have have lengths of $L\approx 4-100$ Mm,
we choose a heating scale height of $s_H=10$ Mm and base heating rates that are
randomly distributed in the logarithmic interval of $E_{H0}=10^{-4},...,10^{-2}$
erg cm$^{-3}$ s$^{-1}$. For the group of long loops with lengths of $L\approx 100-800$ Mm,
we choose near-uniform heating ($s_H=800$ Mm) and volumetric heating rates randomly
distributed in the logarithmic interval of $E_{H0}=0.5\times 10^{-7}, ... ,
0.5\times 10^{-5}$ erg cm$^{-3}$ s$^{-1}$. This choice of heating rates produces a
distribution of loop maximum temperatures (at the loop tops) of $T_e\approx 1-3$
MK, electron densities of $n_e\approx 10^8, ..., 10^{10}$ cm$^{-3}$ at the
footpoints, and $n_e\approx 10^6, ..., 10^{9}$ cm$^{-3}$ at the loop tops.
We show the distribution of loop top temperatures, loop base densities, and
loop top densities in Fig.2. These parameters are considered to be realistic
in the sense that they reproduce typical loop densities and temperatures 
observed with {\sl SoHO} and {\sl TRACE}, as well as correspond to the measured heating
scale heights of $s_H \approx 10-20$ Mm (Aschwanden, Nightingale, \&
Alexander 2000b), for the set of short loops.   

\begin{figure}
\epsfig{file=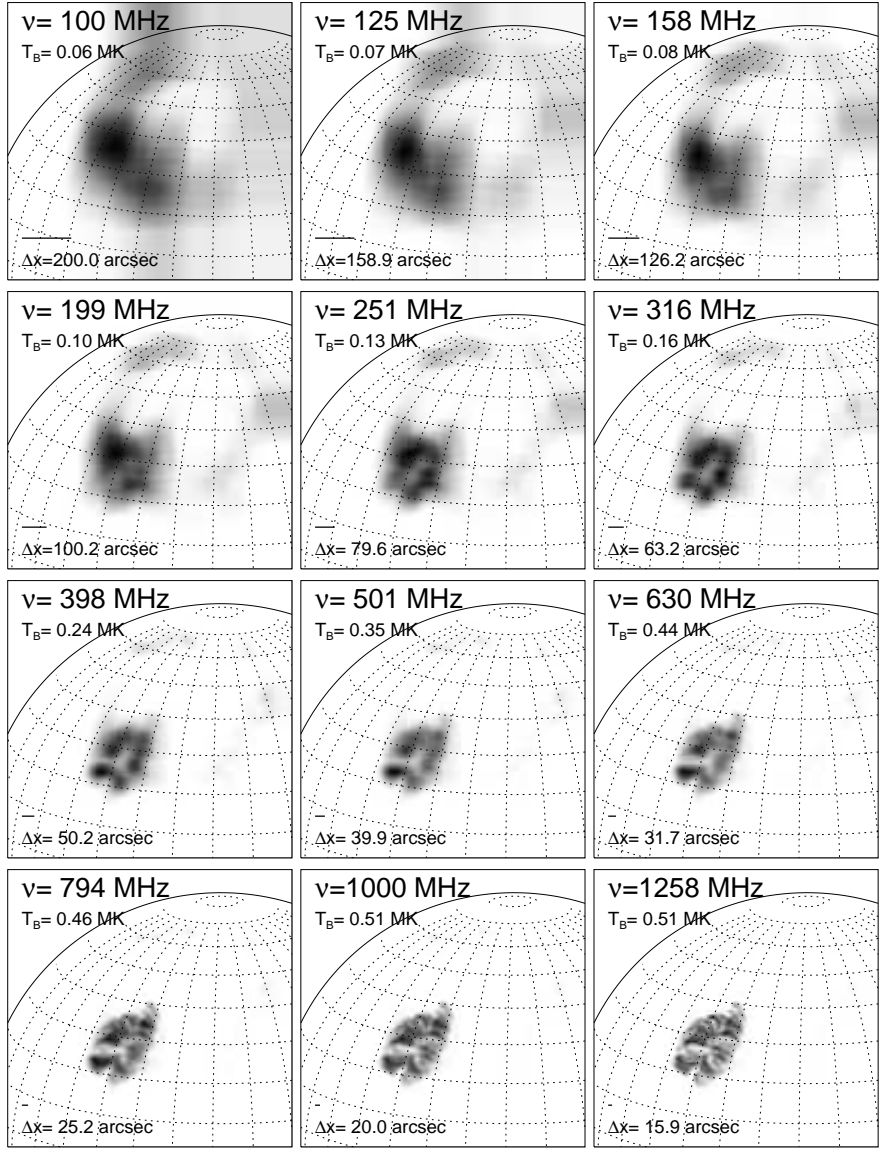,width=\textwidth} 
\caption{Simulation of radio brightness temperature maps of an active
region at 20 frequencies, from $\nu$=100 to 1258 MHz. The maximum
brightness temperature ($T_B$) and the angular resolution $\Delta x$
are indicated in each frame.}
\end{figure}

\begin{figure}
\epsfig{file=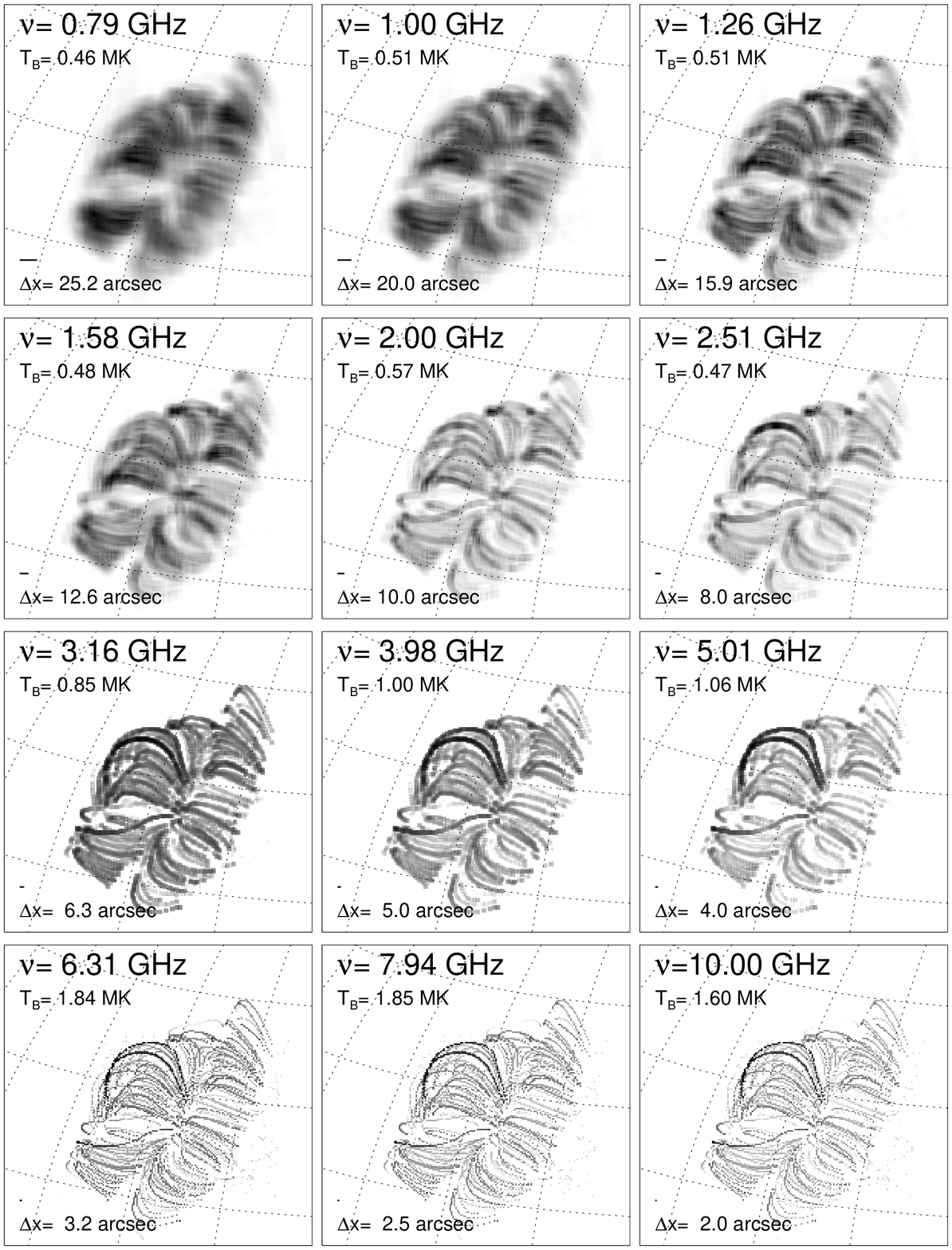,width=\textwidth} 
\caption{Similar representation as in Fig.3, for frequencies of
$\nu$=0.8 to 10 GHz, with a smaller field-of-view than in Fig.3.
The brightness is shown on linear scale in the first two rows,
and on logarithmic scale in the last two rows (with a contrast
of 1:100 in the third row and 1:1000 in forth row).}
\end{figure}

For the simulation of radio images we choose an image size of $512\times 512$
pixels, with a pixel size of 2.25", and 21 frequencies logarithmically
distributed between $\nu=100$ MHz and 10 GHz. To each magnetic field
line we attribute a loop with a width (or column depth) of $w \approx 10^8, ..., 10^9$ cm.
For each voxel, i.e. volume element at ${\bf x}=(x_i,y_j,z_k)$, we calculate the 
free-free absorption coefficient ${\kappa}_{ff}$ (e.g. Lang 1980, p.47),
\begin{equation}
	{\kappa}_{ff}^{\nu}(x_i,y_j,z_k) =
	9.78\times 10^{-3} {n_{e,ijk}^2 \over {\nu}^2 T_{e,ijk}^{3/2}} 
		[24.2+\ln{(T_{e,ijk})}-\ln{(\nu)}] \ ,
\end{equation}
and integrate the opacity ${\tau}_{ff}^{\nu}$ along the line-of-sight $z$,
\begin{equation}
	{\tau}_{ff}^{\nu}(x_i,y_j,z_k) = \int_{-\infty}^{z} {\kappa}_{ff}^{\nu}(x_i,y_j,z_k) dz' \ ,
\end{equation}
to obtain the radio brightness temperature $T_B^{\nu}(x_i,y_j)$ with the radiative
transfer equation (in the Rayleigh-Jeans limit),
\begin{equation}
	T_B^{\nu}(x_i,y_j) = \int_{-\infty}^{+\infty} T_{e,ijk} \exp^{-{\tau}_{ff}^{\nu}((x_i,y_j,z_k)}
		{\kappa}_{ff}^{\nu}(x_i,y_j,z_k) dz \ ,
\end{equation}
The simulated images for the frequency range of $\nu$=100 MHz to 10 GHz are shown
in Figs.3 and 4. The approximate instrumental resolution is rendered by smoothing
the simulated images with a boxcar that corresponds to the instrumental resolution
of FASR,
\begin{equation}
	w_{res} = {20" \over {\nu}_{GHz} } .
\end{equation}

A caveat needs to be made, that the real reconstructed radio images may reach this
theoretical resolution only if a sufficient number of Fourier components are available,
either from a large number of baselines (which scale with the square of the number of
dishes) or from aperture synthesis (which increases the number of Fourier components
during Earth rotation proportionally to the accumulation time interval). Also,
we did not include here the effects of angular scattering due to turbulence or other
coronal inhomogeneities (Bastian 1994, 1995).

\subsection{	Peak Brightness Temperature 	}

\begin{figure}
\epsfig{file=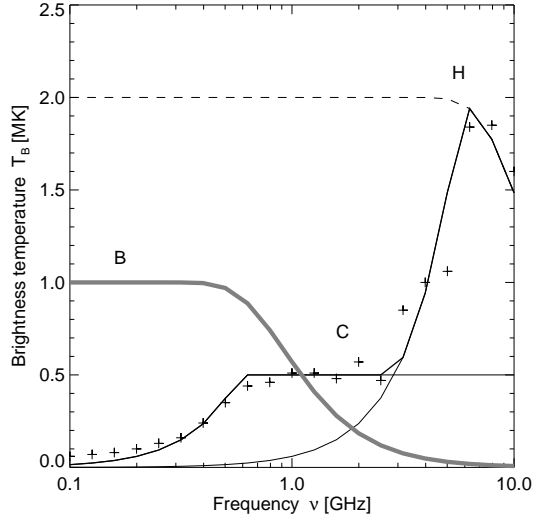,width=\textwidth} 
\caption{The radio peak brightness temperature $T_B$ is shown as function of frequency $\nu$:
for the background corona (B), for cool (C) fat loops (T=0.5 MK, $n_e=10^{11}$ cm$^{-3}$, w=25 Mm),
and hot (H) thin loops (T=2.0 MK, $n_e=10^{11}$ cm$^{-3}$, w=2.5 Mm). The cross symbols
indicate the peak brightness temperatures observed in the simulated maps (Figs.3 and 4),
while the medium-thick line represents the combined model of hot and cool loops. The dashed
line indicates the expected brightness temperature of hot loops if no beam dilution
due to the instrumental angular resolution would occur. The thick grey curve (B) 
indicates a model of the background corona.}
\end{figure}

The intensity of radio maps is usually specified in terms of the observed brightness 
temperature $T_B$. We list the peak brightness temperature in each map in Figs.3 and 4.
We see that a maximum brightness is observed in the second-last map in Fig.4, with
$T_B=1.85$ MK at a frequency of $\nu=7.94$ GHz. Let us obtain some understanding of
the relative brightness temperatures $T_B(\nu )$ as function of frequency $\nu$, in
order to faciliate the interpretation of radio maps. We plot the peak brightness
temperature $T_B(\nu )$ of the simulated maps as function of frequency in Fig.5 
(cross symbols). There are two counter-acting effects that reduce the brightness
temperature: First, the loops become optically thin at high frequencies due to the
${\nu}^{-2}$-dependence of the free-free opacity (Eq.1). Hot loops with a temperature 
of T=2.0 MK, a density of $n_e=10^{11}$ cm$^{-3}$, and a width of w=2.5 Mm are optically
thick below $\nu \lapprox 5$ GHz, so the brightness temperature would match the electron
temperature $T_B=T_e$ (dashed line in Fig.5), but falls off at higher frequencies, i.e.
$T_B(\nu > 5$ GHz)$<T_e$. 

The second effect that reduces the brightness temperature
is the beam dilution, which has a ${\nu}^2$-dependence below the critical frequency
where structures are unresolved. The effectively observed brightness temperature
$T_B^{eff}(\nu )$ due to beam dilution for a structure with width $w$ is 
\begin{equation}
        T_B^{eff}(\nu ) = T_B \times 
        \left\{ \begin{array}{cc}
                \left( {\nu \over {\nu}_{crit} }\right)^2 & {\rm for}\ \nu < {\nu}_{crit}(w)  \\
                1                                  & {\rm for}\ \nu > {\nu}_{crit}(w)
                \end{array}
        \right.
\end{equation}
where the critical frequency ${\nu}_{crit}(w)$ depends on the width $w$ of the structure
and is for FASR according to Eq.(4),
\begin{equation}
	{\nu}_{crit}(w) = { 20" \over w" } \ [GHz] \ .
\end{equation}
Because the brightness drops drastically below ${\nu}_{crit}\approx 5$ GHz in Fig.5,
we conclude that the width of the unresolved structures is about $w"=20"/5=4"=3$ Mm.
Therefore we can understand the peak brightness temperatures in the maps, as shown
in Fig.5 (crosses) in the range of $\nu \approx 3-10$ GHz with a combination
of these two effects of free-free opacity and beam dilution. 

Below a frequency of $\nu \lapprox 3$ GHz, we see that another group of loops contributes
to the peak brightness of the maps. We find that the peak brightness below 3 GHz can 
adequately be understood by a group of cooler loops with a temperature of $T=0.5$ MK,
densities of $n_e=10^{11}$ cm$^{-3}$, and widths of $w=25$ Mm (Fig.5). Thus cool loops
dominate the brightness at low frequencies, and hot loops at higher frequencies.

In the simulations in Figs.3 and 4 we have not included the background corona. 
In order to give a comparison of the effect of the background corona we calculate
the opacity for a space-filling corona with an average temperature of $T=1.0$ MK,
an average density of $n_e=10^9$ cm$^{-3}$, and a vertical (isothermal) scale height of 
$w \approx {\lambda}_T \approx 50$ Mm. The brightness temperature of this background corona is shown
with a thick grey curve (labeled B) in Fig.5. According to this estimate, the background
corona overwhelms the brightest active region loops at frequencies of ${\nu} \lapprox
1$ GHz. From this we conclude that it might be difficult to observe active region
loops at decimetric frequencies ${\nu} \lapprox 1.0$ GHz, unless they are very high
and stick out above a density scale height, i.e. at altitudes of $h \gapprox 50$ Mm.
In conclusion, the contrast of active region loops in our example seems to be best
at frequencies of $\nu \approx 5$ GHz, but drops at both sides of this optimum
frequency (see Fig.5). 

\begin{figure}
\epsfig{file=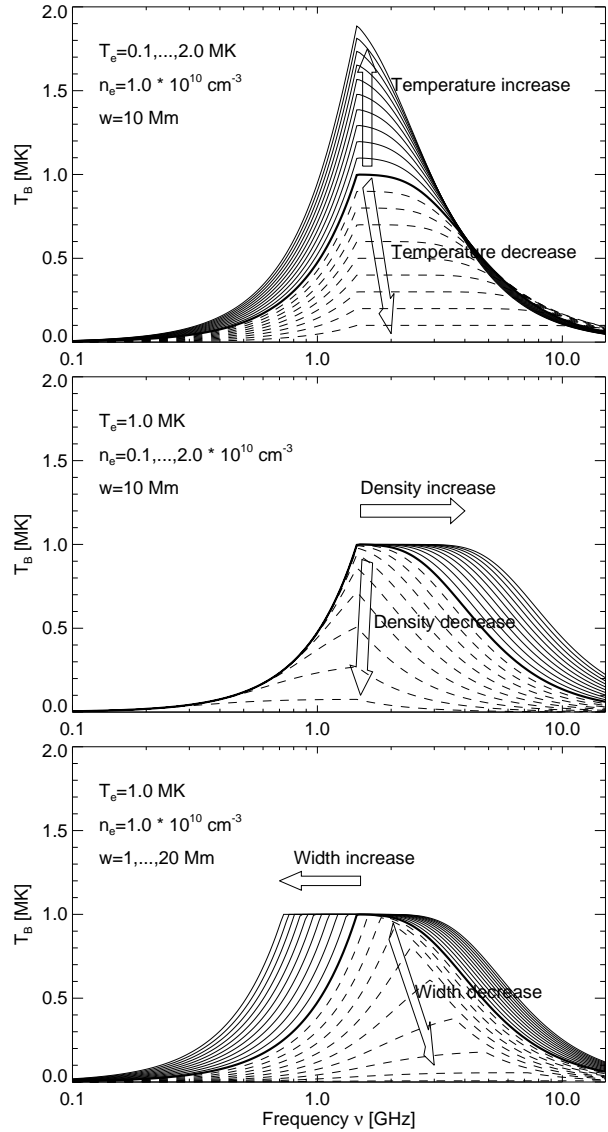,width=\textwidth} 
\caption{The variation of the radio brightness temperature spectrum $T_B(\nu )$ of a loop
by varying the temperature $T_e$ (top panel), the electron density $n_e$ (middle panel), and the 
loop width $w$ (bottom panel). In each of the three panels, one parameter is varied
from 10\%, 20\%, ..., 90\% (dashed curves) to 110\%, 120\%,... , 200\% (solid lines). The
reference curve with parameters $T_e=1.0$ MK, $n_e=10^{10}$ cm$^{-3}$, and $w=10$ Mm
is indicated with a thick line. The arrows indicate the spectral shift of the peak.}
\end{figure}

\subsection{	Temperature and Density Diagnostic of Loops	}

FASR will provide simultaneous sets of images $I(x,y,\nu )$ at many frequencies $\nu$.
In other words, for every image position $(x_i,y_j)$, a spectrum $T_B^{ij}(\nu )$ can
be obtained. A desirable capability is temperature and density diagnostic of active
region loops. Let us parameterize the projected position of a loop by a length
coordinate $s_k$, $k=1,...,n$, e.g. $[x_i=x(s_k),y_j=y(s_k)]$. If we manage to 
determine the temperature $T_e(x_i,y_j)$ and density $n_e(x_i,y_i)$ at every loop position 
$(x_i,y_j)$, we have a diagnostic of the temperature profile $T(s)$ and density
profile $n_e(s)$ of an active region loop. Thus, the question is whether we can
manage to extract a temperature $T_e$ and density $n_e$ from a brightness temperature
spectrum $T_B(\nu )$ at a given pixel position $(i,j)$. In order to illustrate the
feasibility of this task, we show the brightness temperature spectrum $T_B(\nu)$ of
a typical active region loop in Fig.6, and display its variation as a function of
the physical ($T_e$, $n_e$) and geometric ($w$) parameters.    

We define a typical active region loop by an electron temperature $T_e=1.0$ MK,
an electron density $n_e=10^{10}$ cm$^{-3}$, and a width $w=10$ Mm. Such a loop is
brightest at frequencies of $\nu \approx 1.5-3.0$ GHz (Fig.6; thick curve).
The loop is fainter at higher frequencies because free-free emission becomes
optically thin, while it is optically thick at lower frequencies. The reason why
the loop is also fainter at low frequencies is because of the beam dilution
at frequencies where the instrument does not resolve the loop diameter.
If we increase the temperature, the brightness temperature increases, and
vice versa decreases at lower electron temperatures (Fig.6 top). If we
increase the density, the critical frequency where the loop becomes optically
thin shifts to higher frequencies, while the peak brightness temperature
decreases for lower densities (Fig.6, middle panel). If we increase the width
of the loops, the brightness temperature spectrum is bright in a much larger
frequency range, because we shift the critical frequency for beam dilution
towards lower frequencies, while the overall brightness temperature decreases
for a smaller loop width (Fig.6 bottom). Based on this little tutorial, one
can essentially understand how the optimization works in spectral fitting 
(e.g. with a forward-fitting technique) to an observed brightness temperature 
spectrum $T_B(\nu )$.  

\begin{figure}
\epsfig{file=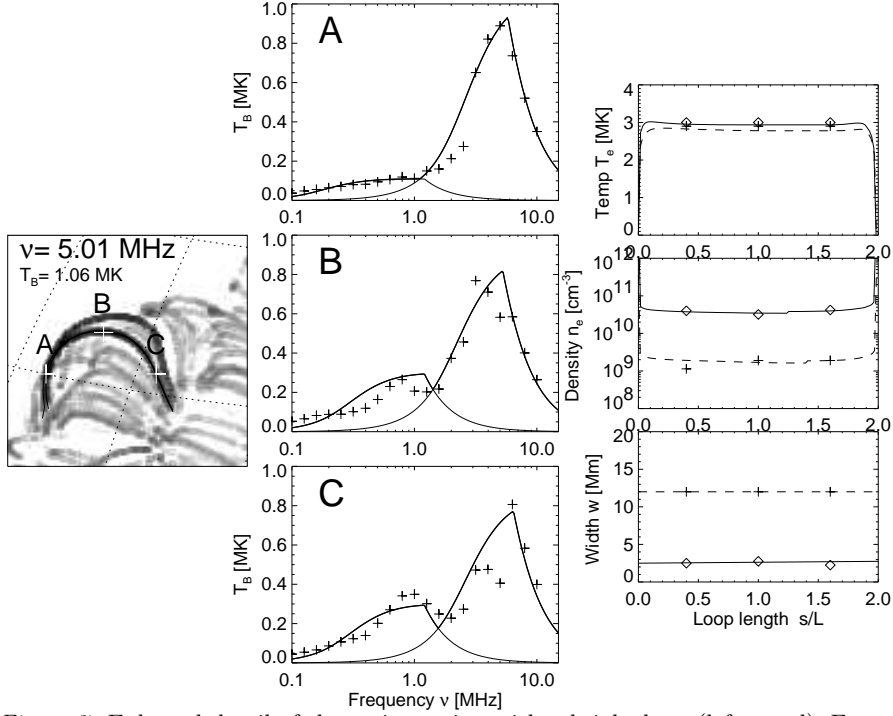,width=\textwidth} 
\caption{Enlarged detail of the active region with a bright loop (left panel).
From the measured brightness temperature spectra $T_B(\nu)$ (crosses in middle
panels) at the three loop locations (A,B,C) we fit theoretical spectra and
determine the temperatures (right top panel), densities (right middle panel),
and loop widths (right bottom panel) at the three loop locations (A,B,C).}
\end{figure}

To demonstrate how the density and temperature diagnostic works in practice,
we pick a bright loop seen at ${\nu}=5.0$ GHz in Fig.4, which we show as an
enlarged detail in Fig.7 (left panel). We pick three locations (A,B,C) along
the loop and extract the brightness temperature spectra $T_B(\nu)$ from the
simulated datacube $T_B(x,y,\nu )$ at the locations (A,B,C), shown in Fig.7
(three middle panels). Each spectrum shows two peaks, which we interpret
as two cospatial loops. For each spectral peak we can therefore roughly
fit a loop model, constrained by three parameters each, i.e. $[T_e, n_e, w]$.
We can now fit a brightness temperature spectrum $T_B^{eff}(\nu )$ to the
observed (or simulated here) spectrum $T_B^{obs}(\nu)$, physically
defined by the same radiation transfer model for free-free emission as
in Eqs.(1-5), but simplified by the approximation of constant parameters
$(T_e, n_e)$, and thus a constant absorption coefficient ${\kappa}_{ff}(\nu)$,
over the relatively small spatial extent of a loop diameter $w$,
\begin{equation}
	{\kappa}_{ff}(\nu) =
	9.78\times 10^{-3} {n_e^2 \over {\nu}^2 T_e^{3/2}} 
		[24.2+\ln{(T_e)}-\ln{(\nu)}] \ ,
\end{equation}
\begin{equation}
	{\tau}_{ff}(\nu) = {\kappa}_{ff}(\nu) \ w \ ,
\end{equation}
\begin{equation}
	T_B(\nu) = T_e \left( 1 - \exp^{-{\tau}_{ff}(\nu)}\right) ,
\end{equation}
\begin{equation}
        T_B^{eff}(\nu ) = T_B \times
        \left\{ \begin{array}{cc}
                \left( {\nu \over {\nu}_{crit} }\right)^2 & {\rm for}\ \nu < {\nu}_{crit}(w)  \\
                1                                  & {\rm for}\ \nu > {\nu}_{crit}(w)
                \end{array}
        \right.
\end{equation}
What can immediately be determined from the observed brightness temperature
spectra $T_B^{obs}(\nu)$ are the frequencies of the spectral peaks 
(Fig.7, middle panels), which are
found around ${\nu}_{peak}=1.2$ and 6.0 GHz. Based on the tutorial given in Fig.6
it is clear that these spectral peaks demarcate the critical frequencies where
structures become unresolved. Thus we can immediately determine the diameters
of the two loops with Eq.(6), i.e. $w_1=20"/1.2=17"=12.0$ Mm and
$w_2=20"/6.0=3.3"=2.4$ Mm. The only thing left to do is to vary the temperature
and density and to fit the model (Eqs.7-10) to the observed spectrum. For an
approximate solution (shown as smooth curves in the middle panels of Fig.7) 
we find $T_1=3.0$ MK and $n_1=4\times 10^{10}$ cm$^{-3}$ for the first loop
(with width $w_1=2.5$ Mm and spectral peak at ${\nu}_1=6.0$ GHz), and 
$T_2=2.9$ MK and $n_2=1.9\times 10^{9}$ cm$^{-3}$ for the second loop
(with width $w_2=12$ Mm and spectral peak at ${\nu}_1=1.2$ GHz). The resulting
temperature $T_e(s)$ and density profiles $n_e(s)$ along the loops are shown
in Fig.7 (right panels). This approximate fit is just an example to illustrate
the concept of forward-fitting to FASR tomographic data. More information can
be extracted from the data by detailed fits with variable loop cross-section
along the loop and proper deconvolution of the projected column depth across 
the loop diameter (which is a function of the aspect angle between the line-of-sight
and the loop axis). For a proper determination of the inclination angle of the
loop plane, the principle of {\sl dynamic stereoscopy} can be applied 
(Aschwanden et al. 1999; see Appendix A therein for coordinate transformations
between the observers reference frame and the loop plane).
Of course, our example is somehow idealized, in practice there will be confusion
by adjacent or intersecting loops, as well as confusion by other radiation
mechanisms, such as gyroresonance emission that competes with free-free emission
at frequencies of $\nu \gapprox 5$ GHz near sunspots. 

\subsection{	Radio versus EUV and soft X-ray diagnostics	} 

We can ask whether temperature and density diagnostic of coronal loops is better
done in other wavelengths, such as in EUV and soft X-rays (e.g. Aschwanden
et al. 1999), rather than with radio tomography. Free-free emission in EUV and
soft X-rays is optically thin, which has the advantage that every loop along
a line-of-sight is visible to some extent, while loops in optically thick plasmas
can be hidden at radio wavelengths. On the other side, the line-of-sight confusion
in optically thin plasmas is larger in EUV and soft X-rays, 
in particular if multiple loops along the
same line-of-sight have similar temperatures. Different loops along a line-of-sight
can only be discriminated in EUV and soft X-rays if they have significantly different 
temperatures, so that they show different responses in lines with different
ionization temperatures. Two cospatial loops that have similar temperatures but
different widths cannot be distinguished by EUV or soft X-ray detectors.
In radio wavelengths, however, even cospatial loops with similar temperatures,
as the two loops in our example in Fig.7 ($T_1=3.0$ MK and $T_2=2.9$ MK), can
be separated if they have different widths. The reason is that they have
different critical frequencies ${\nu}_{crit}(w)$ where they become resolved,
and thus show up as two different peaks in the brightness temperature spectrum
$T_B^{eff}(\nu)$. Radio tomography has therefore a number of unique advantages
over loop analysis in EUV and soft X-ray wavelengths: (1) a ground-based instrument
is much less costly than a space-based instrument, (2) a wide spectral radio wavelength
range (decimetric, centimetric) provides straightforwardly diagnostic over a wide 
temperature range, while an equivalent temperature diagnostic in EUV and soft X-rays
would require a large number of spectral lines and instrumental filters, 
(3) optically thick radio emission is most sensitive to cool plasma, which is
undetectable in EUV and soft X-rays, except for absorption in the case of
very dense cool plasmas, and (4)
radio brightness temperature spectra can discriminate multiple cospatial structures
with identical temperatures based on their spatial scale, which is not possible
with optically thin EUV and soft X-ray emission. 

\section{	CHROMOSPHERIC AND CORONAL MODELING 	}

The vertical density and temperature structure of the chromosphere, transition region,
and corona has been probed in soft X-rays, EUV, and in radio wavelengths, but
detailed models that are consistent in all wavelengths are still unavailable.
Comprehensive coverage of the multi-thermal and inhomogeneous 
solar corona necessarily requires either many wavelength filters 
in soft X-rays and EUV, or many radio frequencies, for which FASR will be 
the optimum instrument.  

We illustrate the concept of how to explore the vertical structure of
the chromosphere and corona with a few simple examples. We know that the
corona is highly inhomogeneous along any line-of-sight, so a 3D model has to be 
composed of a distribution of many magnetic fluxtubes, each one representing
a mini-atmosphere with its own density and temperature structure, 
being isolated from each other due to the low value of the plasma-beta, 
i.e. $\beta = p_{thermal}/p_{magn}=2 n_e k_B T_e/(B^2/8\pi ) \ll 1$. 
The confusion due to inhomogeneous temperatures and densities is largest
for line-of-sights above the limb (due to the longest column depths with
contributing opacity), and is smallest for line-of-sights near the solar 
disk center, where we look down through the atmosphere in vertical direction.
  
The simplest model of the atmosphere is given by the hydrostatic equilibrium 
in the isothermal approximation, $T(h)=const$, where the hydrostatic scale height 
${\lambda}_T$ is proportional to the electron temperature $T$, i.e.
\begin{equation}
    {\lambda}_T = { k_B T \over \mu m_p g_{\odot} } = {\lambda}_0
        \Bigl({T \over 1\ {\rm MK }}\Bigr) \
\end{equation}
with ${\lambda}_0=47$ Mm for coronal conditions, with $\mu m_p$ the average ion
mass (i.e. $\mu \approx 1.3$ for H:He=10:1) and $g_{\odot}$ the solar gravitation.
The height dependence of the electron density is for gravitational pressure balance,
\begin{equation}
        n_e(h) = n_0 \ \exp\bigl[ -{(h - h_0) \over
                         {\lambda}_0 T} \bigr]  \ .
\end{equation}
where $n_0=n(h_0)$ is the base electron density. 
This expression for the density $n_e(h)$ can then be inserted into the free-free 
absorption coefficient ${\kappa}(h,\nu )$, with $T(h)=const$ in the isothermal 
approximation,
\begin{equation}
	{\kappa}_{ff}(h,\nu) =
	9.78\times 10^{-3} {n_e^2(h) \over {\nu}^2 T(h)^{3/2}} 
		[24.2+\ln{T(h)}-\ln{(\nu)}] \ ,
\end{equation}
At disk center, we can set the altitude $h$ equal to the line-of-sight coordinate $z$, so
that the free-free opacity ${\tau}_{ff}(h,\nu)$ integrated along the line-of-sight $h=z$ is,
\begin{equation}
        {\tau}_{ff}(h,\nu) = \int_{-\infty}^{h} {\kappa}_{ff}(h',\nu)\ dh' \ ,
\end{equation}
and the radio brightness temperature $T_B(\nu)$ is then
\begin{equation}
        T_B(\nu) = \int_{-\infty}^{0} T(h) \exp^{-{\tau}_{ff}(h,\nu)}
                {\kappa}_{ff}(h,\nu) dh \ .
\end{equation}
With this simple model we can determine the mean temperature $T(h)$ by fitting
the observed brightness temperature spectra $T_B(\nu)$ to the theoretical
spectra (Eq.15) by varying the temperature $T(h)=const$ (in Eqs.13-14).
The expected brightness temperature spectra for an isothermal corona with
temperatures of $T=1.0$ MK and $T=5.0$ MK and a base density of $n_0=10^9$
cm$^{-3}$ are shown in Fig.8. We see that the corona becomes 
optically thin ($T_B \ll T_e$) at frequencies of ${\nu}\gapprox 1-2$ GHz 
in this temperature range that is typical for the Quiet Sun.

\begin{figure}
\epsfig{file=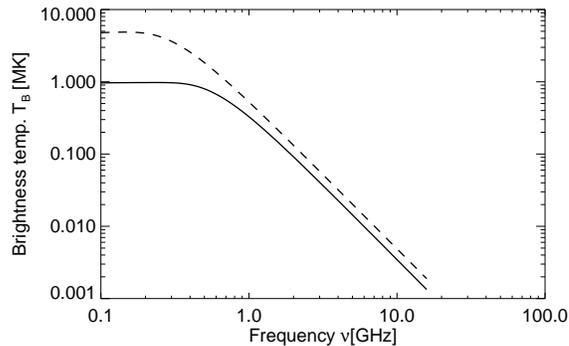,width=\textwidth} 
\caption{Quiet Sun brightness temperature spectrum for an isothermal corona
with $T=1.0$ MK (solid line) or $T=5.0$ MK (dashed line) with a base density
of $n_0=10^9$ cm$^{-3}$.}
\end{figure}

\begin{figure}
\epsfig{file=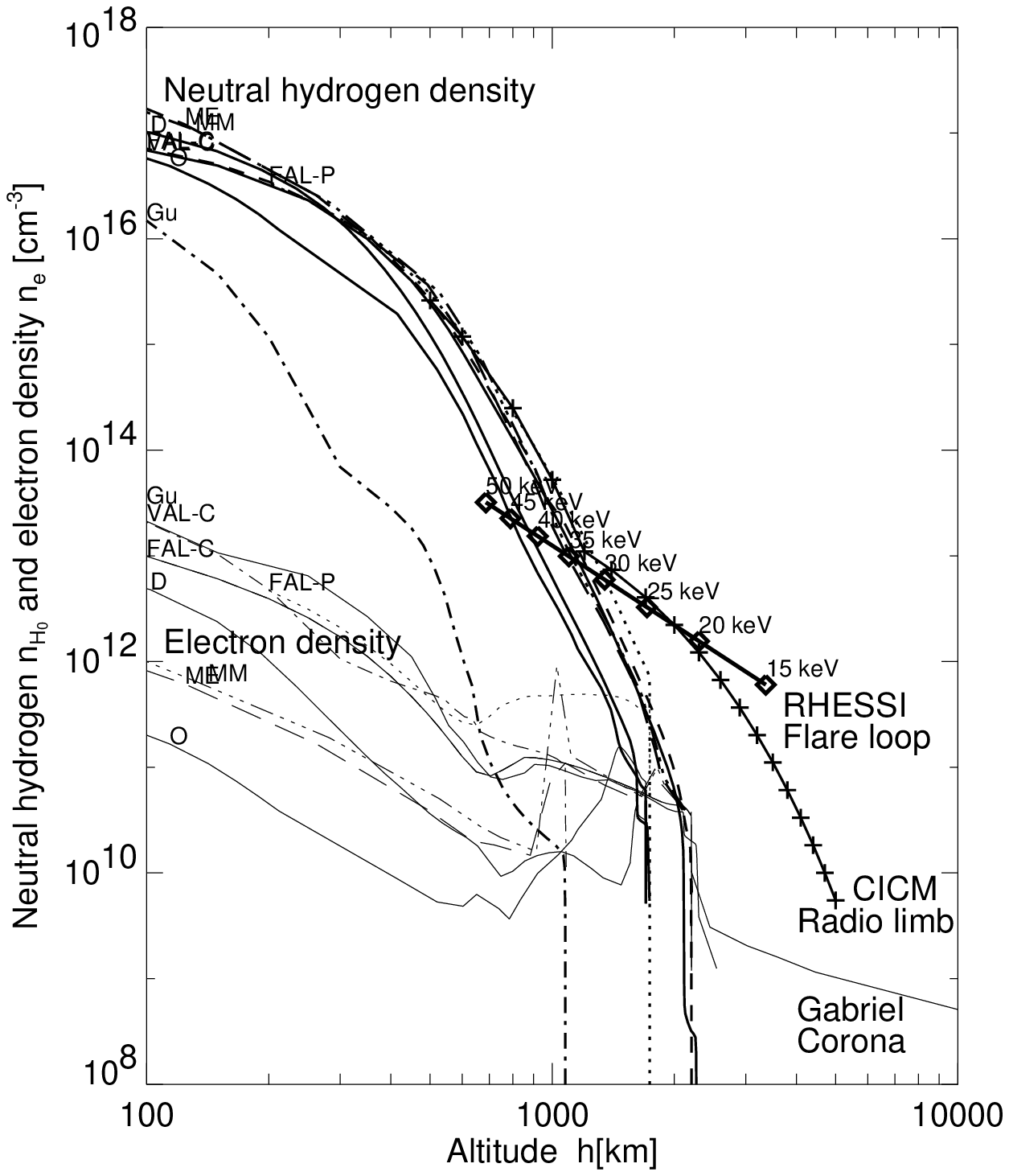,width=\textwidth} 
\caption{
A compilation of chromospheric and coronal density models:
VAL-C = Vernazza, Avrett, \& Loeser (1981), model C;
FAL-C = Fontenla, Avrett, \& Loeser (1990), model C;
FAL-P = Fontenla, Avrett, \& Loeser (1990), model P;
G = Gu, Jefferies et al. (1997); MM = Maltby
et al., (1986), model M; ME = Maltby et al., (1986), model E;
D  = Ding \& Fang (1989); O  = Obridko \& Staude (1988);
Gabriel = Gabriel (1976), coronal model;
CICM = Caltech Irreference Chromospheric Model, radio sub-millimeter
limb observations (Ewell et al. 1993), RHESSI flare loop 
(Aschwanden, Brown, \& Kontar 2002).}
\end{figure}

These hydrostatic models in the
lower corona, however, have been criticized because of the presence
of dynamic phenomena, such as spicu\-lae, which may contribute to an
extended chromosphere in the statistical average. The spicular
extension of this dynamic chromosphere has been probed with
high-resolution measurements of the {\sl Normal Incidence
X-Ray Telescope (NIXT)} (Daw, DeLuca, \& Golub 1995) as well as with radio
submillimeter observations during a total eclipse (Ewell et al.
1993). Using the
radio limb height measurements at various mm and sub-mm wavelengths
in the range of 200-3000 $\mu$m (Roellig et al. 1991; Horne et al.
1981; Wannier et al. 1983; Belkora et al. 1992; Ewell et al. 1993),
an empirical {\sl Caltech Irreference Chromospheric Model
(CICM)} was established, which fits the observed limb heights between
500 km and 5000 km in a temperature regime of $T=4410$ K to $T=7500$
K (Ewell et al. 1993), shown in Fig.9.  We see that these radio limb
measurements yield electron densities that are 1-2 orders
of magnitude higher in the height range of 500-5000 km than predicted
by hydrostatic models (VAL, FAL, Gabriel 1976), which was
interpreted in terms of the dynamic nature of spiculae
(Ewell et al. 1993). This enhanced density in the extended chromosphere 
has also been corroborated with recent {\sl RHESSI} measurements
(Fig.8; Aschwanden, Brown, \& Kontar 2002). Hard X-rays
mainly probe the total neutral and ionized hydrogen density
that governs the bremsstrahlung and the total bound
and free electron density in collisional energy losses,
while the electron density $n_e(h)$ inferred from the radio-based
measurements is based on free-free emission, and shows a remarkably
good agreement in the height range of $h\approx 1000-3000$ km. 
The extended chromosphere produces
substantially more opacity at microwave frequencies than 
hydrostatic models (e.g. Gabriel 1976).  

\begin{figure}
\epsfig{file=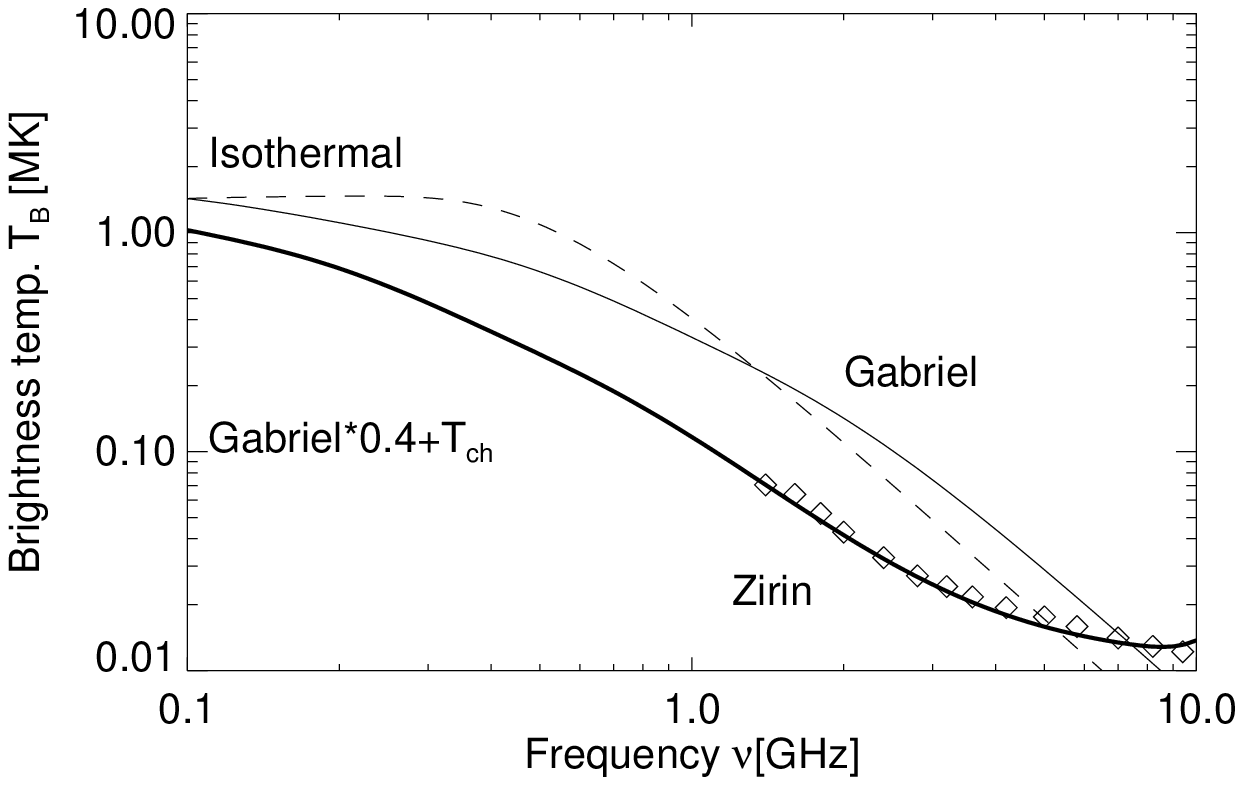,width=\textwidth} 
\caption{Quiet Sun brightness temperature spectrum for an isothermal corona
with $T=1.5$ MK with a base density of $n_0=10^9$ cm$^{-3}$ (solid thin
line), for the coronal model by Gabriel (1976) (thin solid line), and
for a modified Gabriel model (thick solid line).}
\end{figure}

The atmospheric structure thus needs to be explored with more general 
parameterizations of the density $n_e(h)$ and temperature $T_e(h)$ 
structure than hydrostatic models provide. For instance, each 
of the atmospheric models shown in Fig.9 provides different
functions $n_e(h)$ and $T_e(h)$. Observational tests of these models
can simply be made by forward-fitting of the parameterized
height-dependent density $n_e(h)$ and temperature profiles $T_e(h)$, 
using the expressions for free-free emission (Eqs.13-15).   
In Fig.10 we illustrate this with an example. The datapoints (shown
as diamonds in Fig.10) represent radio observations of the solar
limb at frequencies of ${\nu}=1.4-18$ GHz during the solar minimum
in 1986-87 by Zirin, Baumert, \& Hurford (1991).
We show in Fig.10 an isothermal hydrostatic model for a coronal 
temperature of $T_e=1.5$ MK and a base density of $n_e=10^{9}$ cm$^{-3}$,
as well as the hydrostatic model of Gabriel (1976),
of which the density profile $n_e(h)$ is shown in Fig.9. The
Gabriel model was calculated based on the expansion of the magnetic
field of coronal flux tubes over the area of a supergranule (canopy
geometry). The geometric expansion factor and the densities at the 
lower boundary in the transition region (given by the chromospheric
VAL and FAL models, see Fig.9) then constrains the coronal density
model $n_e(h)$, which falls off exponentially with height in an isothermal
fluxtube in hydrostatic equilibrium. We see that the Gabriel model
roughly matches the isothermal hydrostatic model (see Fig.10), but 
does not exactly match the observations by Zirin et al. (1991). 
However, if we multiply the Gabriel model by a factor of 0.4,
to adjust for solar cycle minimum conditions, and add a temperature
of $T_e=11,000$ K to account for an optically thick chromosphere
(similar to the values determined by Bastian, Dulk, and Leblanc 1996),
we find a reasonably good fit to the observations of Zirin (thick curve
in Fig.10). This example demonstrates that radio spectra in the
frequency range of ${\nu} \approx 1-10$ GHz are quite sensitive to
probe the physical structure of the chromosphere and transition region. 

\section{ 	FUTURE FASR SCIENCE 		}

With our study we illustrated some basic applications of frequency
tomography as can be expected from FASR data. We demonstrated how
physical parameters from coronal loops in active regions, from
the Quiet-Sun corona, and from the chromosphere and transition
region can be retrieved. Based on these capabilities we expect
that the following science goals can be efficiently studied with
future FASR data:

\begin{enumerate}

\item{  The electron density $n_e(s)$ and electron temperature profile 
	$T_e(s)$ of individual active region loops can be retrieved, which
	constrain the heating function $E_H(s)$ along the loop in 
	the momentum and energy balance hydrodynamic equations. This
	enables us to test whether a loop is in hydrostatic equilibrium
	or evolves in a dynamic manner. Detailed dynamic studies
	of the time-dependent heating function $E_H(s,t)$ may reveal
	the time scales of intermittent plasma heating processes, which
	can be used to constrain whether AC or DC heating processes
	control energy dissipation. Ultimately, such quantitative
	studies will lead to the determination and identification of
	the so far unknown physical heating mechanisms, a long-thought
	goal of the so-called {\sl coronal heating problem}. Radio
	diagnostic is most sensitive to cool dense plasma, but is
	also sensitive continuously up to the highest temperatures, and
	this way nicely complements EUV and soft X-ray diagnostic.}

\item{	Because coronal loops are direct tracers of closed coronal magnetic
	field lines, the reconstruction of the 3D geometry of loops, as
	it can be mapped out with multi-frequency data from FASR in a
	tomographic manner, this information can be used to test
	theoretical models based on magnetic field extrapolations 
	from the photosphere. The circular polarization of free-free
	emission contains additional information on the magnetic field
	(Grebinskij et al. 2000; Gelfreikh 2002; Brosius 2002), 
	as well as gyroresonance emission provides 
	direct measurements of the magnetic field by its proportionality
	to the gyrofrequency (Lee et al. 1998; White 2002; Ryabov 2002). 
	Ultimately, such studies may constrain the non-potentiality and the
	localization of currents in the corona.}

\item{	The density $n_e(h)$ and temperature profile $T_e(h)$ of the 
	chromosphere, transition region, and corona can be determined
	in the Quiet Sun from brightness temperature spectra $T_B(\nu)$, 
	with least confusion at disk center. Parameterized models of
	the density and temperature structure, additionally constrained 
	by the hydrodynamic equations and differential emission measure
	distributions, can be forward-fitted to the observed radio
	brightness temperature spectra $T_B(\nu )$. This provides a
	new tool to probe physical conditions in the transition
	region, deviations from hydrostatic equilibria, and diagnostic of
	dynamic processes (flows, turbulence, waves, heating, cooling) 
	in this little understood interface to the corona.}  

\item{	Since free-free emission is most sensitive to cool dense plasma,
	FASR data will also be very suitable to study the origin,
	evolution, destabilization, and eruption of filaments, which
	seem to play a crucial role in triggering and onset of {\sl coronal
	mass ejections} (Vourlidas 2002). Ultimately, the information to 
	forecast CMEs may be chiefly exploited from the early evolution 
	of filaments.}

\end{enumerate}
Previous studies with multi-frequency instruments (VLA, OVRO, Nan\-\c{c}ay, RATAN-600)
allowed only crude attempts to pioneer tomographic 3D-modeling of the solar
corona, because of the limitations of a relatively small number of Fourier
components and a sparse number of frequencies. FASR will be the optimum instrument
to faciliate 3D diagnostic of the solar corona on a routine basis, which is
likely to lead to groundbreaking discoveries in long-standing problems of
coronal plasma physics.

\end{article}

\begin{thebibliography}{}

\newcommand{\ApJ}{\sl Astrophys.J.\ }
\newcommand{\ApJS}{\sl Astrophys.J.Suppl.Ser.\ }
\newcommand{\AaA}{\sl Astron.Astrophys.\ }
\newcommand{\AaAS}{\sl Astron.Astrophys.Suppl.Ser.\ }
\newcommand{\JGR}{\sl J.Geophys.Res.\ }
\newcommand{\SP}{\sl Solar Phys.\ }

\bibitem[]{}{Altschuler,M.D. 1979, in Image Reconstruction from Projections, 
	(ed. G.T.Herman, Berlin:Springer, p.105}
\bibitem[]{}{Aschwanden,M.J. and Bastian,T.S. 1994a, \ApJ {\bf 426}, 425}
\bibitem[]{}{Aschwanden,M.J. and Bastian,T.S. 1994b, \ApJ {\bf 426}, 434}
\bibitem[]{}{Aschwanden,M.J., Lim,J., Gary,D.E., and Klimchuk,J.A. 1995, \ApJ {\bf 454}, 512}
\bibitem[]{}{Aschwanden,M.J., Newmark,J.S., Delaboudiniere,J.P., Neupert,W.M., 
	Klimchuk,J.A., Gary,G.A., Portier-Fornazzi,F., and Zucker,A. 1999, \ApJ {\bf 515}, 842}
\bibitem[]{}{Aschwanden,M.J., Alexander,D., Hurlburt,N., Newmark,J.S., Neupert,W.M., 
	Klimchuk,J.A., and G.A.Gary 2000a, \ApJ {\bf 531}, 1129}
\bibitem[]{}{Aschwanden,M.J., Nightingale,R.W., and Alexander,D. 2000b, \ApJ {\bf 541}, 1059}
\bibitem[]{}{Aschwanden,M.J. and Schrijver,K.J. 2002, \ApJS {\bf 142}, 269}
\bibitem[]{}{Aschwanden,M.J., Brown,J.C., and Kontar,E.P. 2002, \SP (in press)}
\bibitem[]{}{Bastian,T.S. 1994, \ApJ {\bf 426}, 774}
\bibitem[]{}{Bastian,T.S. 1995, \ApJ {\bf 439}, 494}
\bibitem[]{}{Bastian,T.S., Dulk,G.A., and Leblanc,Y. 1996, \ApJ {\bf 473}, 539}
\bibitem[]{}{Batchelor,D.A. 1994, \SP {\bf 155}, 57}
\bibitem[]{}{Belkora,L., Hurford,G.J., Gary,D.E. and Woody,D.P. 1992, \ApJ {\bf 400}, 692}
\bibitem[]{}{Berton.R. and Sakurai,T. 1985, \SP {\bf 96}, 93}
\bibitem[]{}{Bogod,V.M., and Grebinskij,A.S. 1997, \SP {\bf 176}, 67}
\bibitem[]{}{Brosius,J.W. 2002, (in this volume)}
\bibitem[]{}{Davila,J.M. 1994, \ApJ {\bf 423}, 871}
\bibitem[]{}{Daw,A., DeLuca,E.E., and Golub,L. 1995, \ApJ {\bf 453}, 929}
\bibitem[]{}{Ding, M.D. and Fang, C. 1989, \AaA {\bf 225}, 204.}
\bibitem[]{}{Ewell, M.W.Jr., Zirin, H., Jensen, J.B., and Bastian, T.S.
        1993, \ApJ {\bf 403}, 426.}
\bibitem[]{}{Fontenla, J.M., Avrett, E.H., and Loeser, R. 1990, \ApJ {\bf 355}, 700.}
\bibitem[]{}{Gabriel, A.H. 1976, {\sl Royal Society (London), Philosophical
        Transactions, Series A}, {\bf 281}, no. 1304, p. 339.}
\bibitem[]{}{Gary,A., Davis,J.M., and Moore,R. 1998, \SP {\bf 183}, 45}
\bibitem[]{}{Gelfreikh,G.B. 2002, (in this volume)}
\bibitem[]{}{Grebinskij,A., Bogod,V., Gelfreikh,G., Urpo,S., Pohjolainen,S. 
	and Shibasaki,K. 2000, \AaAS {\bf 144}, 169}
\bibitem[]{}{Gu, Y., Jefferies, J.T., Lindsey, C. and Avrett, E.H. 1997, \ApJ {\bf 484}, 960.}
\bibitem[]{}{Horne, K., Hurford, G.J., Zirin, H., and DeGraauw, Th. 1981,
        \ApJ {\bf 244}, 340.}
\bibitem[]{}{Hurlburt,N.E., Martens,P,C.H., Slater,G.L., and Jaffey,S.M.
	1994, in {\sl Solar Active Region Evolution: Comparing Models with
	Observations}, ASP Conf. Ser. {\bf 68}, 30}
\bibitem[]{}{Lang,K.R. 1980, {\sl Astrophysical Formulae. A Compendium
	for the Physicist and Astrophysicist}, Berlin: Springer}
\bibitem[]{}{Lee,J.W., McClymont,A.N., Mikic,Z., White,S.M., and Kundu,M.R. 1998, \ApJ {\bf 501}, 853}
\bibitem[]{}{Koutchmy,S., and Molodensky,M.M. 1992, {\sl Nature} {\bf 360}, 717}
\bibitem[]{}{Koutchmy,S., Merzlyakov, V. L., and Molodensky, M. M. 2001, 
	{\sl Astronomy Reports} {\bf 45}/10, 834}
\bibitem[]{}{Liewer,P.C., Hall,J.R., DeJong,M., Socker,D.G., Howard,R.A., 
	Crane,P.C., Reiser,P., Rich,N., Vourlidas,A. 2001, \JGR {\bf 106}/A8, 15903}
\bibitem[]{}{Maltby, P., Avrett, E.H., Carlsson, M., Kjeldseth-Moe, O.,
        Kurucz, R.L., and Loeser, R. 1986, \ApJ {\bf 306}, 284.}
\bibitem[]{}{Nitta,N., VanDriel-Gestelyi,L., and Harra-Murnion,L,K. 1999, \SP {\bf 189}, 181}
\bibitem[]{}{Obridko, V.N. and Staude, J. 1988, \AaA {\bf 189}, 232.}
\bibitem[]{}{Roellig, T.L., Becklin, E.E., Jefferies, J.T., Kopp, G.A.,
        Lindsey, C.A., Orral, F.Q., and Werner, M.W., 1991, \ApJ {\bf 381}, 288.}
\bibitem[]{}{Ryabov,V.B. 2002, (in this volume)}
\bibitem[]{}{Vernazza, J.E., Avrett, E.H., and Loeser, R. 1981, \ApJS {\bf 45}, 635.}
\bibitem[]{}{Vourlidas,A. 2002, (in this volume)}
\bibitem[]{}{Wannier, P.G., Hurford, G.J., and Seielstad, G.A. 1983, \ApJ {\bf 264}, 660.}
\bibitem[]{}{White,S.M. 2002, (in this volume)}
\bibitem[]{}{Zidowitz,S. 1999, \JGR {\bf 104}/A5, 9727}
\bibitem[]{}{Zirin,H., Baumert,B.M., and Hurford,G.J. 1991, \ApJ {\bf 370}, 779}

\end{thebibliography}
\end{document}